\documentclass[12pt]{article}

\title{Maximally Entangled Two-Qutrit Quantum Information States and De Gua's Theorem for Tetrahedron}

\author{Oktay K Pashaev \\
Department of Mathematics \\ Izmir Institute of Technology \\ Izmir 35430, Turkey}

\begin{document}

\maketitle              

\begin{abstract}
   Geometric relations between separable and entangled two-qubit and two-qutrit quantum information states are studied. To characterize entanglement of two qubit states, we establish a relation between reduced density  matrix and the concurrence. For the rebit states, the geometrical meaning of concurrence as double area of a parallelogram is found and for generic qubit states it is expressed by determinant of the complex Hermitian inner product metric, where reduced density matrix  coincides with the inner product metric.
In the case of generic two-qutrit state,
 for reduced density matrix we find Pythagoras type relation,
where the concurrence  is expressed by sum 
of all $2 \times 2$ minors of $3\times3$ complex matrix.
 For maximally entangled two-retrit state, this relation is just
De Gua's theorem or a three-dimensional analog of the Pythagorean theorem for triorthogonal tetrahedron areas.
Generalizations of our results for arbitrary two-qudit states are discussed

\end{abstract}



 %
 \section{Introduction}
  Here we consider the qubit, qutrit and generic qudit units of quantum information, determined by position system with bases 2, 3, and arbitrary $d$, correspondingly.
\subsubsection{Qubit}
Each qubit state $|\psi \rangle \in {\cal H} = {\cal C}^2$ can be represented as a linear combination of $|0\rangle$ and $|1\rangle$ states, 
\begin{equation}
|\psi\rangle=c_0|0\rangle+c_1|1\rangle,
\end{equation}
where 
\begin{equation}|0\rangle=\left(\begin{array}{c}
1\\0\end{array}\right), \,\,\,\,\,|1\rangle=\left(\begin{array}{c}
0\\1\end{array}\right),
\end{equation}
and $ |c_0|^2 + |c_1|^2 = 1 $.             

\subsubsection{Qutrit}

Each qutrit state $|\psi \rangle \in {\cal H} = {\cal C}^3$ can be represented as a linear combination 
\begin{equation}
|\psi\rangle=c_0|0\rangle+c_1|1\rangle+c_2|2\rangle,
\end{equation}
of three computational states
$|0\rangle$, $|1\rangle$ and $|2\rangle$ states, 
where 
\begin{equation}
|0\rangle=\left(\begin{array}{c}
1\\0\\0\end{array}\right), \,\,\,\,|1\rangle=\left(\begin{array}{c}
0\\1\\0\end{array}\right), \,\,\,\,|2\rangle=\left(\begin{array}{c}
0\\0\\1\end{array}\right)
\end{equation}
and
$$ |c_0|^2 + |c_1|^2  + |c_{2}|^2 = 1 .             $$

\subsubsection{Qudit}

Each qutdit state $|\psi \rangle \in {\cal H} = {\cal C}^d$ is a linear combination 
\begin{equation}
|\psi\rangle=c_0|0\rangle+c_1|1\rangle+\cdots+c_{d-1}|d-1\rangle,
\end{equation}
of states $|0\rangle$, $|1\rangle$,..., $|d-1\rangle$, 
where 
\begin{equation}
|0\rangle=\left(\begin{array}{c}
1\\ \vdots\\0\end{array}\right), \,\,\,\ldots, \,\,\,\,|d-1\rangle=\left(\begin{array}{c}
0\\ \vdots\\1\end{array}\right)
\end{equation}

and
$$ |c_0|^2 + |c_1|^2 + ... + |c_{d-1}|^2 = 1 .             $$

\section{Separability and Entanglement of Two Qubit States }

The generic two qubit state 
\begin{equation}
|\psi\rangle = c_{00}|00\rangle+c_{01}|01\rangle+c_{10}|10\rangle+c_{11}|11\rangle, \label{2qubit}
\end{equation}
where
$
|c_{00}|^2+|c_{01}|^2+  |c_{10}|^2        + |c_{11}|^2=1, 
$
can be represented in following form
\begin{equation}
|\psi\rangle=|0\rangle|\phi\rangle+|1\rangle|\xi\rangle, \label{form}
\end{equation}
 where the one qubit states are
$
|\phi\rangle = c_{00}|0\rangle+c_{01}|1\rangle, \label{phi}$, 
$|\xi\rangle = c_{10}|0\rangle+c_{11}|1\rangle\label{xi}
$.

 Two qubit state (\ref{2qubit}) is separable if and only if states $|\phi\rangle$ and $|\xi\rangle $ are linearly dependent.

\subsubsection{Concurrence and Area}

For the rebit states 
$
|\phi\rangle = r_{00}|0\rangle+r_{01}|1\rangle, \label{phi}$, 
$|\xi\rangle = r_{10}|0\rangle+r_{11}|1\rangle\label{xi}
$,
with real coefficients $(r_{00}, r_{01}) \equiv \vec r_0$ and $(r_{10}, r_{11}) \equiv \vec r_1$ the linear dependence of these two dimensional vectors can be expressed by the area formula in plane.

Two vectors $\vec r_0$ and  $\vec r_1$ in plane  determine parallelogram with area
\begin{equation}
A = |\vec r_0 \times \vec r_1| = \left|\det \left|
\begin{array}{cc}
r_{00} & r_{01}\\
r_{10} & r_{11}
\end{array} \right|\right|.
\end{equation}
The normalization condition 
\begin{equation}
\vec r_0^2 + \vec r_1^2 =1 \label{normalization}
\end{equation}
restricts this area by maximal value $A_{max} = \frac{1}{2}$, so that Eq. (\ref{normalization}) becomes the Pythagoras theorem for the orthogonal vectors, corresponding to rectangle, inscribed to the circle with radius $\frac{1}{2}$.

The concurrence 
\begin{equation}  C=2
\left|
\begin{array}{cc}
r_{00} & r_{01}\\
r_{10} & r_{11}
\end{array}
\right|=2A
\end{equation}
is bounded function
$$0\leq C\leq1 .$$

\subsubsection{Concurrence and Reduced Density Matrix} 

The concurrence, as a pure geometrical quantity is connected  with physical characteristics of entangled two qubit states, represented by reduced density matrix \cite{PP}.
 
For generic two qubit state, the trace of squared reduced density matrix $\rho_A$ and the concurrence $C$
are connected by "Pythagoras theorem"
\begin{equation}
tr \, \rho^2_A + \frac{1}{2} C^2 =1.
\end{equation}

If $C =0$, then $tr \, \rho^2_A =1$ and the state is separable.
If $C \neq 0$, then $tr \, \rho^2_A < 1$ and the state is entangled (non-separable).

\section{Separability and Entanglement of Two Qutrit States}

For generic  two qutrit state 
\begin{equation}
|\psi\rangle=\sum_{i,j=0,1,2}c_{ij}|ij\rangle,
\end{equation}
the density matrix is
\begin{equation}
\rho=|\psi\rangle\langle\psi|=\sum_{i,j,i',j'}c_{ij}\overline{c_{i'j'}}|ij\rangle\langle i'j'|.
\end{equation}

The reduced density matrix $\rho_A = tr_B \, \rho$ for the generic two qutrit state is
\begin{equation}
\rho_A=\sum_{i,i'}(c_{i0}\overline{c_{i'0}}|i\rangle\langle i'|+c_{i1}\overline{c_{i'1}}|i\rangle\langle i'|+c_{i2}\overline{c_{i'2}}|i\rangle\langle i'|)
\end{equation}
or in matrix form $\rho_A=$
$$\left(\begin{array}{ccc}
|c_{00}|^2+|c_{01}|^2+|c_{02}|^2 \quad&\quad c_{00} \bar c_{10}+ c_{01}\bar c_{11}+c_{02}\bar c_{12} \quad&\quad c_{00}\bar c_{20}+c_{01}\bar c_{21}+c_{02}\bar c_{22} \\\\
\bar c_{00} c_{10}+ \bar  c_{01} c_{11}+ \bar c_{02} c_{12}  \quad&\quad |c_{10}|^2+|c_{11}|^2+|c_{12}|^2 \quad&\quad c_{10}\bar c_{20}+c_{11}\bar c_{21}+c_{12}\bar c_{22} \\\\
\bar c_{00} c_{20}+ \bar c_{01} c_{21}+ \bar c_{02} c_{22} \quad&\quad \bar c_{10} c_{20}+ \bar c_{11} c_{21}+ \bar c_{12} c_{22} \quad&\quad |c_{20}|^2+|c_{21}|^2+|c_{22}|^2\end{array}\right)$$
 and due to normalization
$$|c_{00}|^2+|c_{10}|^2+|c_{20}|^2+|c_{01}|^2+|c_{011}|^2+|c_{21}|^2+|c_{02}|^2+|c_{12}|^2+|c_{22}|^2=1,$$
$tr \,\rho_A =1$.

To decide if the reduced density matrix corresponds to pure or mixed state, we calculate $tr \, \rho^2_A$.

Trace of the squared reduced density matrix satisfies the "Phytagoras theorem"
\begin{equation}
tr \rho^2_A + \frac{1}{2} {C}^2 =1,\label{PT3}
\end{equation}
where the square of the qutrit concurrence is defined by sum
\begin{eqnarray}
\frac{1}{4} C^2 = \left|\left|\begin{array}{cc} c_{00} & c_{01} \\ c_{10} & c_{11} \end{array} \right|\right|^2 + \left|\left|\begin{array}{cc} c_{00} & c_{02} \\ c_{10} & c_{12} \end{array} \right|\right|^2 + \left|\left|\begin{array}{cc} c_{01} & c_{02} \\ c_{11} & c_{12} \end{array} \right|\right|^2 \label{M}\\
 +\left|\left|\begin{array}{cc} c_{00} & c_{01} \\ c_{20} & c_{21} \end{array} \right|\right|^2 + \left|\left|\begin{array}{cc} c_{00} & c_{02} \\ c_{20} & c_{22} \end{array} \right|\right|^2 + \left|\left|\begin{array}{cc} c_{01} & c_{02} \\ c_{21} & c_{22} \end{array} \right|\right|^2 \nonumber \\
 +\left|\left|\begin{array}{cc} c_{10} & c_{11} \\ c_{20} & c_{21} \end{array} \right|\right|^2 + \left|\left|\begin{array}{cc} c_{10} & c_{12} \\ c_{20} & c_{22} \end{array} \right|\right|^2 + \left|\left|\begin{array}{cc} c_{11} & c_{12} \\ c_{21} & c_{22} \end{array} \right|\right|^2 \nonumber
\end{eqnarray}
or
\begin{equation}{C}^2 = 4 \sum_{*}     \left|\left|\begin{array}{cc} c_{**} & c_{**} \\ c_{**} & c_{**} \end{array} \right|\right|^2,   \label{C2}                                     \end{equation}
which includes modulus squares of all $2\times2$ minors of $3 \times 3$ matrix $(C)_{ij} = c_{ij} $.

The concurrence for two qutrit states is defined as 
\begin{equation}
{C} = 2 \sqrt{\sum_{*}     \left|\left|\begin{array}{cc} c_{**} & c_{**} \\ c_{**} & c_{**} \end{array} \right|\right|^2 }.\label{qutritC}
\end{equation}

If $C =0$, then $tr \, \rho^2_A =1$ and the state is separable.
If $C \neq 0$, then $tr \, \rho^2_A < 1$ and the state is entangled (non-separable).

For separable two-qutrit states the matrix $c_{ij} = c_i d_j$ have to be rank one. In this case all $2\times2$ minors are zero and   ${ C} = 0$, so that
$tr \rho^2_A = 1$ and the reduced state is the pure state.

If rank of matrix $C$ is bigger than one, then exists at least  one non-vanishing $2\times2$ minor, so that ${C} \neq 0$ and as follows $tr \rho^2_A < 1$. The reduced state in this case becomes mixed. 

The value of positive number ${\cal C}$ shows how far is given state from the pure one and so, how much it is entangled. The maximally entangled state then should corresponds to
maximal value of ${\cal C}$. 

\section{Geometrical Representation of Two-Retrit State in 3D Space}

Here we examine generic qutrit  states with real coefficients $c_i = r_i$, $c_{ij} = r_{ij}$, $i,j =0,1,2$, and call them the retrit states.
The generic two-retrit state is characterized by three, the one-retrit states $|\phi_0\rangle$, $|\phi_1\rangle$, $|\phi_2\rangle$,
\begin{equation}
|\psi \rangle = |0\rangle |\phi_0\rangle + |1\rangle |\phi_1\rangle + |2\rangle |\phi_2\rangle, 
\end{equation}
with real coordinates.
These coordinates determine corresponding three vectors in 3D Euclidean space:
\begin{equation} {\vec r_0} = (r_{00}, r_{01}, r_{02}),\,\,\, {\vec r_1} = (r_{10}, r_{11}, r_{12})    ,\,\,\, {\vec r_2} = (r_{20}, r_{21}, r_{22}) . \label{vectors} \end{equation}
By taking modulus of the cross product of two vectors,
$$   |{\vec r_i}\times {\vec r_j}|   = A_{ij}                         $$
we find that it is equal to the area $A_{ij}$ of corresponding parallelogram. For our three vectors we have three areas  $A_{01}, A_{02}, A_{12}$  and corresponding squares
 \begin{equation}  |{\vec r_0}\times {\vec r_1}|^2   = A^2_{01} =  \left|\left|\begin{array}{cc} r_{00} & r_{01} \\ r_{10} & r_{11} \end{array} \right|\right|^2 + \left|\left|\begin{array}{cc} r_{00} & r_{02} \\ r_{10} & r_{12} \end{array} \right|\right|^2 + \left|\left|\begin{array}{cc} r_{01} & r_{02} \\ r_{11} & r_{12} \end{array} \right|\right|^2,
 \end{equation}
 \begin{equation}  |{\vec r_0}\times {\vec r_2}|^2   = A^2_{02} =  \left|\left|\begin{array}{cc} r_{00} & r_{01} \\ r_{20} & r_{21} \end{array} \right|\right|^2 + \left|\left|\begin{array}{cc} r_{00} & r_{02} \\ r_{20} & r_{22} \end{array} \right|\right|^2 + \left|\left|\begin{array}{cc} r_{01} & r_{02} \\ r_{21} & r_{22} \end{array} \right|\right|^2,
 \end{equation}
\begin{equation}  |{\vec r_1}\times {\vec r_2}|^2   = A^2_{12} =  \left|\left|\begin{array}{cc} r_{10} & r_{11} \\ r_{20} & r_{21} \end{array} \right|\right|^2 + \left|\left|\begin{array}{cc} r_{10} & r_{12} \\ r_{20} & r_{22} \end{array} \right|\right|^2 + \left|\left|\begin{array}{cc} r_{11} & r_{12} \\ r_{21} & r_{22} \end{array} \right|\right|^2.
 \end{equation}

\subsubsection{Maximally Entangled Retrit States}

Let us denote
\begin{equation}
 A^2=A_{01}^2+A_{02}^2+A_{12}^2, 
\end{equation}
so that the concurrence is
 \begin{equation}{C} = 2 A= 2 \sqrt{A_{01}^2+A_{02}^2+A_{12}^2}.\end{equation}
Substituting to  (\ref{PT3}) this gives formula
 \begin{equation}  tr\, \rho^2_A +  2  (A_{01}^2+A_{02}^2+A_{12}^2)= tr \,\rho^2_A +  2 A^2  = 1.
\end{equation}
 The formula shows that for pure state, when $tr\, \rho^2_A =1$, $A = 0$  and as follows $A_{01} = A_{02} = A_{12} = 0 $. This means  that mutual cross products between all three vectors ${\vec r_0}, {\vec r_1}, {\vec r_2}$ are zero, and
 vectors are directed along the same line. If at least one of the cross products is non vanishing, then $A^2 \neq 0$, hence $tr\, \rho^2_A <1$  and   the state is mixed.
 The mixture increases with increasing value of $A$ and reaches maximal value for maximal $A$.

The maximal value of $A^2 = \frac{1}{3}$ corresponds to
$r_0^2 = r_1^2 = r_2^2 = \frac{1}{3}$.
Then,  for maximally reduced mixed state  
 \begin{equation} tr \rho^2_A = \frac{1}{3} \end{equation}
and maximal concurrence for maximally entangled two-retrit state is
\begin{equation}
C = \frac{2}{\sqrt{3}}.
\end{equation}

 The areas of parallelograms are maximal for orthogonal vectors, so that
$A_{01}^2=r_{0}^2r_{1}^2, 
A_{02}^2=r_{0}^2r_{2}^2,
A_{12}^2=r_{1}^2r_{2}^2$.
   Then
 \begin{equation}   A^2=A_{01}^2+A_{02}^2+A_{12}^2 = r^2_0 r^2_1   + r^2_0 r^2_2 + r^2_1 r^2_2.           \end{equation}
 Applying this formula to normalization condition  and its square
 \begin{eqnarray}
r_{0}^2+r_{1}^2+r_{2}^2=1 ,\\
r_{0}^4+r_{1}^4+r_{2}^4+2( r_{0}^2 r_{1}^2+r_{0}^2 r_{2}^2+r_{1}^2 r_{2}^2 )=1,
\end{eqnarray}
 we get relations
$$ r_{0}^4+r_{1}^4+r_{2}^4+2A^2=1,$$
 $$     A^2= r^2_0 r^2_1   + r^2_0 r^2_2 + r^2_1 r^2_2 .            $$
By using  $
r_{2}^2=1-r_{0}^2-r_{1}^2
$ we can  exclude $r_2$ from the last equation 
so that
 $$ f(r_0, r_1) \equiv A^2 = r_0^2 + r_1^2 - r_0^4 - r_1^4 - r_0^2r_1^2, $$
where
$|r_0|\leq1$ and $|r_1|\leq1$.
By denoting $x \equiv r_0^2$, $y \equiv r_1^2$ the problem reduces 
to find exact extrim values of function
$$ F(x,y) \equiv x + y -x^2 -y^2 -x y.$$ These are solutions of the system
 \begin{eqnarray}
 F_x = 1 - 2x - y = 0,\,\,\,\,
 F_y = 1 - 2y - x = 0.
 \end{eqnarray}
The solution is $x = y = \frac{1}{3}$ and
 the maximal value of $A^2$, corresponding to
 $ r_0^2 = r_1^2 = r_2^2 = \frac{1}{3}                        $
 is equal
 $ A^2_{max} = \frac{1}{3}.$

\subsubsection{Trirectangular Tetrahedron}

 Three vectors ${\vec r_0}, {\vec r_1}, {\vec r_2}$ from origin determines a tetrahedron in 3D Euclidean space.
 Areas of his faces at origin are
 $$ A_{\triangle 01} = \frac{1}{2}A_{01},\,\,\,A_{\triangle 02} = \frac{1}{2}A_{02},\,\,\,A_{\triangle 12} = \frac{1}{2}A_{12}.$$
To calculate area of the face, opposite to the origin we have
 $$A_{\triangle} = \frac{1}{2}A = \frac{1}{2} \left|(\overrightarrow{r_{0}}- \overrightarrow{r_{2}})\times (\overrightarrow{r_{1}}- \overrightarrow{r_{2}}) \right|.$$
 Then by using formula
 \begin{equation}
( \overrightarrow{A} \times \overrightarrow{B} )\cdot( \overrightarrow{C} \times \overrightarrow{D} )=( \overrightarrow{A}\cdot\overrightarrow{C} )( \overrightarrow{B}\cdot\overrightarrow{D})-(\overrightarrow{A}\cdot\overrightarrow{D})(\overrightarrow{B}\cdot\overrightarrow{C})
\end{equation}
we get
 \begin{eqnarray}
A^2&=&( \overrightarrow{r_{0}}-\overrightarrow{r_{2}} )^2( \overrightarrow{r_{1}}-\overrightarrow{r_{2}} )^2-( (\overrightarrow{r_{0}}-\overrightarrow{r_{2}}) \cdot(\overrightarrow{r_{1}}-\overrightarrow{r_{2}}))^2 \nonumber\\
&=&(r_{0}^2+r_{2}^2-2\overrightarrow{r_{0}} \cdot\overrightarrow{r_{2}} )( r_{1}^2+r_{2}^2-2\overrightarrow{r_{1}} \cdot\overrightarrow{r_{2}} )-(\overrightarrow{r_{0}} \cdot\overrightarrow{r_{1}}  - \overrightarrow{r_{0}}\cdot \overrightarrow{r_{2}} - \overrightarrow{r_{2}} \cdot\overrightarrow{r_{1}} +  {r_{2}}^2)^2 .\nonumber\\
\end{eqnarray} 
As we have seen, maximally mixed state corresponds to the orthogonal vectors and  trirectangular tetrahedron.
 A trirectangular tetrahedron is a tetrahedron, where all three face angles at one vertex are right angles.
In this case all scalar products between orthogonal vectors are zero and the area formula becomes
\begin{eqnarray}
A^2 =
=r_{0}^2 r_{1}^2+r_{0}^2 r_{2}^2+r_{1}^2 r_{2}^2, \nonumber
\end{eqnarray} 
 and
 \begin{eqnarray}
A_{\triangle}^2
=\frac{1}{4}(r_{0}^2 r_{1}^2+r_{0}^2 r_{2}^2+r_{1}^2 r_{2}^2). \label{trianglearea}
\end{eqnarray} 
For  areas of faces at origin we have
 $$  A^2_{01}   =   |{\vec r_0}\times {\vec r_1}|^2  =   r_{0}^2 r_{1}^2             $$
  $$  A^2_{02}  =         |{\vec r_0}\times {\vec r_2}|^2  = r_{0}^2 r_{2}^2  $$
  $$  A^2_{12}  =    |{\vec r_1}\times {\vec r_2}|^2    =  r_{1}^2 r_{2}^2          $$
so that the relation between areas is
\begin{equation}A^2 = A^2_{01} +  A^2_{02} + A^2_{12}. \label{area}\end{equation}

\subsection{De Gua's Generalization of Pythagoras Theorem for Areas}

De Gua's theorem is a three-dimensional analog of the Pythagoras theorem or generalization the Pythagoras theorem to a tetrahedron \cite{DC}.
For the trirectangular tetrahedron, the square of the area of the face, opposite to the right-angle corner is the sum of the squares of the areas of the other three faces:
 \begin{equation}    A_{\triangle}^2=A_{\triangle 01}^2+A_{\triangle 02}^2+A_{\triangle 12}^2. \label{degua} \end{equation}
This result follows easily from equations (\ref{trianglearea}) and (\ref{area}) for maximally mixed state, since
$$
A_{\triangle 01} = A_{01} =\frac{1}{2}|r_0||r_1| ,\quad A_{\triangle 02}= A_{02}= \frac{1}{2}|r_0||r_2| \quad A_{\triangle 12}= A_{12}= \frac{1}{2}|r_1||r_2| . \nonumber
$$
This way we have established the relation between maximally entangled two-retrit state and De Gua's theorem or generalization of the Pythagoras theorem to a tetrahedron.

\section{Qutrit Entanglement}

Now we work with generic complex two qutrit state, which can be decomposed
\begin{equation}
|\psi\rangle = |0\rangle |c_0 \rangle + |1\rangle |c_1 \rangle + |2\rangle |c_2 \rangle,\label{qutritstate}
\end{equation}
in terms of one qutrit states
\begin{eqnarray}
|c_0\rangle &=& c_{00} |0\rangle + c_{01} |1\rangle + c_{02} |2\rangle, \\
|c_1\rangle &= &c_{10} |0\rangle + c_{11} |1\rangle + c_{12} |2\rangle,\\
|c_0\rangle& = &c_{20} |0\rangle + c_{21} |1\rangle + c_{22} |2\rangle .
\end{eqnarray}

The external (cross) product of two qutrit states $|c_a\rangle$ and $|c_b\rangle$, ($a,b=0,1,2$) is a qutrit state, denoted as      $|c_a \times c_b \rangle$ and defined by formula
\begin{equation}
|c_a \times c_b \rangle = \sum^2_{i,j,k =0} \epsilon_{ijk} c_{a j} c_{b k} |i\rangle.
\end{equation}
It can be represented as
\begin{equation}
|c_a \times c_b \rangle =
\left|\begin{array}{ccc} |0\rangle & |1\rangle & |2\rangle \\ c_{a0} & c_{a1} & c_{a2}\\ c_{b0} & c_{b1} & c_{b2}\end{array} \right| = 
\left| \begin{array}{cc} c_{a1} & c_{a2} \\ c_{b1} & c_{b2} \end{array} \right| |0\rangle
-\left| \begin{array}{cc} c_{a0} & c_{a2} \\ c_{b0} & c_{b2} \end{array} \right| |1\rangle
+\left| \begin{array}{cc} c_{a0} & c_{a1} \\ c_{b0} & c_{b1} \end{array} \right| |2\rangle.
\end{equation}

By using cross product of states, the concurrence (\ref{qutritC}) can be represented as
\begin{equation}
C = 2 \sqrt{\langle c_0 \times c_1 |c_0 \times c_1 \rangle + \langle c_2 \times c_0 |c_2 \times c_0 \rangle + \langle c_1 \times c_2 |c_1 \times c_2 \rangle } \label{Ccrossform}
\end{equation}

The norms of the cross product states  are
\begin{eqnarray}
 \langle c_0 \times c_1 |c_0 \times c_1 \rangle &=& \left|\left|\begin{array}{cc} c_{00} & c_{01} \\ c_{10} & c_{11} \end{array} \right|\right|^2 + \left|\left|\begin{array}{cc} c_{00} & c_{02} \\ c_{10} & c_{12} \end{array} \right|\right|^2 + \left|\left|\begin{array}{cc} c_{01} & c_{02} \\ c_{11} & c_{12} \end{array} \right|\right|^2,\label{MM}\\
 \langle c_2 \times c_0 |c_2 \times c_0 \rangle &=& \left|\left|\begin{array}{cc} c_{00} & c_{01} \\ c_{20} & c_{21} \end{array} \right|\right|^2 + \left|\left|\begin{array}{cc} c_{00} & c_{02} \\ c_{20} & c_{22} \end{array} \right|\right|^2 + \left|\left|\begin{array}{cc} c_{01} & c_{02} \\ c_{21} & c_{22} \end{array} \right|\right|^2, \nonumber \\
 \langle c_1 \times c_2 |c_1 \times c_2 \rangle &=& \left|\left|\begin{array}{cc} c_{10} & c_{11} \\ c_{20} & c_{21} \end{array} \right|\right|^2 + \left|\left|\begin{array}{cc} c_{10} & c_{12} \\ c_{20} & c_{22} \end{array} \right|\right|^2 + \left|\left|\begin{array}{cc} c_{11} & c_{12} \\ c_{21} & c_{22} \end{array} \right|\right|^2, \nonumber
\end{eqnarray}
and after substituting to (\ref{M}) we get the concurrence formula (\ref{Ccrossform}).

The inner product of two cross product states can be represented as
\begin{equation}
\langle A\times B | C \times D \rangle = \langle A|C\rangle \langle B|D\rangle -
\langle A|D\rangle \langle B|C\rangle
\end{equation}

Norms of cross product states in (\ref{Ccrossform}) are equal
\begin{eqnarray}
\langle c_0\times c_1 | c_0\times c_1 \rangle &=& \langle c_0|c_0\rangle \langle c_1|c_1\rangle -
|\langle c_0|c_1\rangle|^2 ,\label{01}\\ 
\langle c_2\times c_0 | c_2\times c_0 \rangle &=& \langle c_0|c_0\rangle \langle c_2|c_2\rangle -
|\langle c_2|c_0\rangle|^2 ,\label{02}\\ 
\langle c_1\times c_2 | c_1\times c_2 \rangle &=& \langle c_1|c_1\rangle \langle c_2|c_2\rangle -
|\langle c_1|c_2\rangle|^2 .\label{12}
\end{eqnarray}

For maximally entangled two qutrit states the concurrence is
\begin{equation}
C = 2 \sqrt{\langle c_0| c_0 \rangle  + \langle c_1 | c_1 \rangle  + \langle c_2 | c_2 \rangle }.\label{maxC}
\end{equation}

The Hermitian inner product determines the Hermitian metrics
\begin{equation}
g_{ij} = \langle c_i | c_j \rangle = \overline{ \langle c_j | c_i \rangle} = g^+_{ij}.
\end{equation}
By unitary transformation of one qutrit states
$
|\tilde c_i\rangle = U_{ij} |c_j\rangle
$
this metric can be diagonilized, so that all mutual inner products vanish
$\langle \tilde c_0| \tilde c_1 \rangle = \langle \tilde c_0| \tilde c_2 \rangle = \langle \tilde c_1| \tilde c_2 \rangle =0$,
and the metric is $\tilde g_{ij} = diag(\langle \tilde c_0| \tilde c_0 \rangle, \langle \tilde c_1| \tilde c_1 \rangle, \langle \tilde c_2| \tilde c_2 \rangle)$. 
In this case, the concurrence (\ref{Ccrossform}), due to (\ref{01})-(\ref{12}), reduces to the form (\ref{maxC}).

The norms of states for diagonal metric are constrained only by normalization condition
\begin{equation}
tr\, g = \langle \tilde c_0| \tilde c_0 \rangle + \langle \tilde c_1| \tilde c_1 \rangle + \langle \tilde c_2| \tilde c_2 \rangle =1.
\end{equation}
By denoting these norms as $\langle \tilde c_0| \tilde c_0 \rangle = r^2_0$, 
$\langle \tilde c_1| \tilde c_1 \rangle = r^2_1$, $\langle \tilde c_2| \tilde c_2 \rangle = r^2_2$, the problem is reduced to the one we have studied above for the two retrit states. Thus, we have the following result.

The concurrence for maximally entangled two-qutrit state is
\begin{equation}
C = \frac{2}{\sqrt{3}}.
\end{equation}

\subsubsection{Example} 

As an example we consider two-qutrit state 

\begin{equation}
|\psi \rangle = \frac{1}{\sqrt{3}}(|00\rangle + |11\rangle + |22\rangle).
\end{equation}
It is maximally entangled state with concurrence $C = 2/\sqrt{3}$. Due to real coefficients, this state is the retrit state and De Gua's theorem (\ref{degua}) is given by relation
\begin{equation}
\frac{1}{12} = \frac{1}{36} + \frac{1}{36} + \frac{1}{36},
\end{equation}
for corresponding trirectangular tetrahedron, with three equal sides $1/\sqrt{3}$.

\paragraph{Notes and Comments.}
The geometrical characterization of entangled states by concurrence and area, allowed us to establish intriguing link between the key property of quantum states in quantum information theory, to be maximally entangled and
3-dimensional extension of Pythagoras theorem. 
We like to note  that "Pythagoras theorem" for entanglement (\ref{C2})
can be generalized to arbitrary two qudit states. The generic two qudit state is determined by $d \times d$ complex matrix from coefficients $c_{ij}$, ($i,j=0, 1,...,d-1$). Then by calculating the reduced density matrix we get following equation
\begin{equation}
1 - tr \, \rho^2_A = 2 \sum^{d-1}_{i<i'} \sum^{d-1}_{j<j'} \left|\begin{array}{cc} c_{ij} & c_{ij'} \\ c_{i'j} & c_{i'j'} \end{array}\right|^2,
\end{equation}
where summation includes modulus square of all $2\times2$ minors of $d \times d$ matrix $c_{ij}$. This allows us to define the concurrence for entanglement of two qudit states.
Relation between maximally entangled two qudit states and generalizations of Pythagoras theorem to higher dimensions is under the study.
This work was supporting by BAP project 2022IYTE-1-0002.

%
%

\end{document}